\documentclass{mem}
\usepackage{natbib}\usepackage{txfonts}\usepackage{balance}
\usepackage{graphicx}
\usepackage[a4paper]{hyperref}
\idline{1}{1}
\begin{document}
\def\teff{$T\rm_{eff }$}
\def\kms{$\mathrm {km s}^{-1}$}

\title{
The evolution of star forming galaxies with the Wide Field X-ray Telescope
}

   \subtitle{}

\author{
P. \,Ranalli\inst{1} 
          }

  \offprints{P. Ranalli}

\institute{
Universit\`a di Bologna -- Dipartimento di Astronomia,
via Ranzani 1, 40127 Bologna, Italy
\email{piero.ranalli@oabo.inaf.it}
}

\authorrunning{Ranalli }

\titlerunning{star forming galaxies with the WFXT}

\abstract{ Star forming galaxies represent a small yet sizable
  fraction of the X-ray sky ($1\%$--$20\%$, depending on the flux).
  X-ray surveys allow to derive their luminosity function and
  evolution, free from uncertainties due to absorption. However, much
  care must be put in the selection criteria to build samples clean
  from contamination by AGN. Here we review the possibilities offered
  by the proposed WFXT mission for their study. We analyze the
  expected luminosity and redshift distributions of star forming
  galaxies in the proposed WFXT surveys. We discuss the impact of such
  a mission on the knowledge of the cosmic star formation history, and
  provide a few suggestions.

}
\maketitle{}

%
%

\def\S{Sect.}
\newcommand{\Hii}{\ion{H}{ii} }
\newcommand{\Log}{\mathrm{Log}}
\newcommand{\XMM}{XMM-{\em Newton}}
\newcommand{\xmm}{XMM-{\em Newton}}
\newcommand{\chandra}{{\em Chandra}}
\newcommand{\spitzer}{{\em Spitzer}}
\newcommand{\flux}{{\sc Flux} }
\newcommand{\lum}{{\sc Lum.} }
\newcommand{\lognlogs}{Log~$N$--Log~$S$}
\newcommand{\FIR}{{\rm FIR} }
\newcommand{\SFR}{{\rm SFR} }
\newcommand{\LX}{L_{\rm X} }
\newcommand{\Lsx}{L_{0.5-2} }
\newcommand{\Lhx}{L_{2-10} }
\newcommand{\de}{{\rm d}}
\newcommand{\ergs}{erg s$^{-1}$}
\newcommand{\ergscmq}{erg s$^{-1}$ cm$^{-2}$}
\newcommand{\ergscmqdegq}{erg s$^{-1}$ cm$^{-2}$ deg$^{-2}$}
\newcommand{\ergsHz}{erg s$^{-1}$ Hz$^{-1}$}
\newcommand{\nWmqsr}{nW m$^{-2}$ sr$^{-1}$}
\newcommand{\e}[1]{\cdot 10^{#1}}
\newcommand{\figuraacolori}{{\em (In colour only in the electronic edition)}\/\ }
\newcommand{\fxott}{\Log (F_{{\rm X-ray}}/F_{{\rm opt}})}
\newcommand{\etalum}{{\eta_{\rm l}}}
\newcommand{\etaden}{{\eta_{\rm d}}}

\newcommand{\wide}{\textit{wide}}
\newcommand{\medium}{\textit{medium}}
\newcommand{\deep}{\textit{deep}}

\section{Introduction}
\label{sect_norman}

The X-ray luminosity of star forming galaxies (SFG; they usually are
spiral galaxies without AGN activity) appears to be a reliable,
absorption-free estimator of star formation
\citep{rcs03}.  This is justified on the basis that the X-ray
luminosities are linearly and tightly correlated with the radio and
FIR ones, which in turn are commonly used as star formation rate (SFR)
indicators. Thus, the
X-ray emission of SFG may be considered as a tool to investigate the
cosmic star formation history. To this end, the study of the X-ray
luminosity function (XLF) of galaxies and of its evolution represents
a necessary step. \citet[][hereafter RCS05]{rcs05} built a local
($z=0$) XLF of SFG and investigated the possibilities for
evolution. In this paper, we build on the RCS05 XLF and methods to
explore the possible contribution of the Wide Field X-ray Telescope
(WFXT) mission to our understanding of the SFG content of the
universe, by analyzing the expected luminosity and redshift
distributions.

The WFXT is a proposed mission which aims to perform very wide and
moderately deep X-ray surveys. By taking a different approach to
mirror design than the classical Wolter type-1
\citep{wide-field-optics}, it could achieve a very large field of view
($\sim 1$ deg$^2$) while maintaining a good angular resolution ($\sim
5\arcsec$) and a large effective area ($\sim 1$ m$^2$) in the 0.1-7
keV band \citep{wfxt}.

\begin{figure}[tp]
  \centering
   \includegraphics[width=.95\columnwidth]{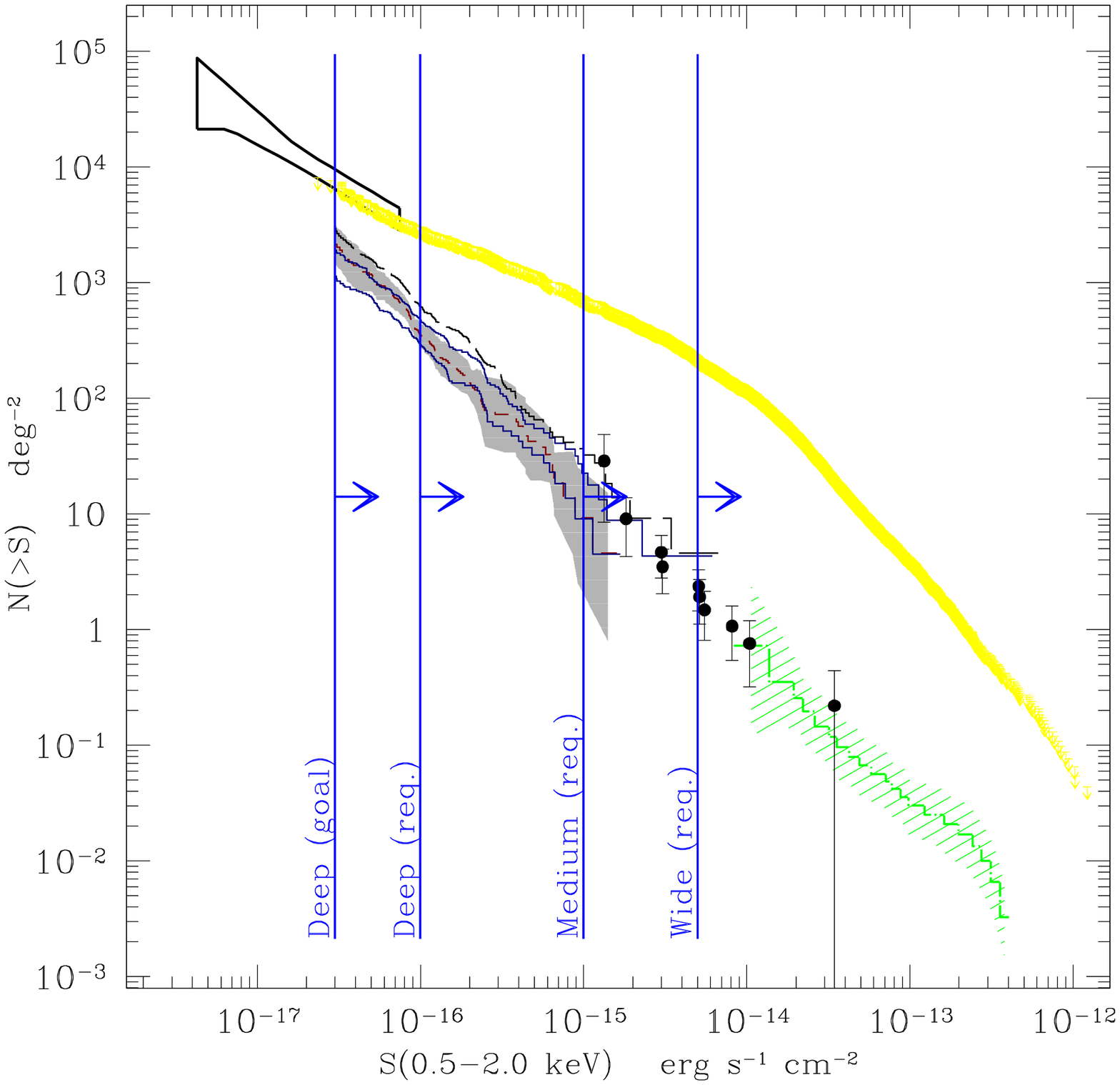}
  \caption{
    Observed X-ray number counts in today's surveys, and planned WFXT
    limiting fluxes. The thick
    upper line and the horn show the observed \lognlogs\ 
    for all X-ray sources in the \chandra\ Deep Fields
    \citep{moretti03} and the limits from the fluctuation analysis
    \citep{miyaji02a}. The bundle of histograms and data
    points shows several determinations of the star-forming galaxies
    \lognlogs\ (see text). 
    The vertical
    lines illustrate the  limiting fluxes of the planned surveys.
  }
  \label{fig:conteggi}
\end{figure}

\begin{table}[tp]
  \caption{Covered area (deg$^2$) and limiting fluxes (\ergscmq\ in
    the 0.5--2.0 keV band)
    of the proposed WFXT surveys.}
  \centering
  \begin{tabular}{lrrr}
                        &wide     &medium     &deep  \\
   \hline
   area                 &20000    &3000       &100 \\
   flux (req.)   &$5\e{-15}$ &$10^{-15}$  &$10^{-16}$ \\
   flux (goal)          &$3\e{-15}$ &$5\e{-16}$ &$3\e{-17}$ 
  \end{tabular}
  \label{tab:areaflux}
\end{table}

Such a telescope would be able to observe a number of X-ray sources
far exceeding all those known today. While X-ray surveys mainly detect
AGN, star-forming galaxies (SFG) are also present, comprising a
fraction in the range 1\%--20\% (depending on the flux) of all the
sources detected in the 0.5--2.0 keV band.  Three major surveys are
envisaged with the WFXT, covering different amounts of the sky at
different limiting fluxes and named \wide, \medium\ and \deep\
(Fig~\ref{fig:conteggi}).  Their limiting soft X-ray fluxes correspond
broadly to those probed in ROSAT \citep{tajer05}, \xmm\
\citep{georgakakis04} and deep \chandra\
\citep[][RCS05]{bauer04,norman04} surveys of SFG.  In
Fig.~\ref{fig:conteggi} we show the total \lognlogs\ from X-ray
surveys, and different estimates of the SFG number counts.

Depending on technological developments, both a {\em requirement} and
a {\em goal} value for the limiting fluxes can be quoted. Reaching the
goals could extend the number of detected objects by a factor $\sim
5$. However, given the early stage of the mission, here we will
consider only the requirements, and regard the goal flux limit for the
\deep\ survey only.

We assume $H_0=70$ km s$^{-1}$ Mpc$^{-1}$,
$\Omega_{\rm M}=0.3$ and $\Omega_{\Lambda}=0.7$.

\section{The LF and evolution of star-forming galaxies}
\label{sect_FIR2XLF}

\begin{figure*}[tp]    
   \centering
   \includegraphics[width=.46\textwidth]{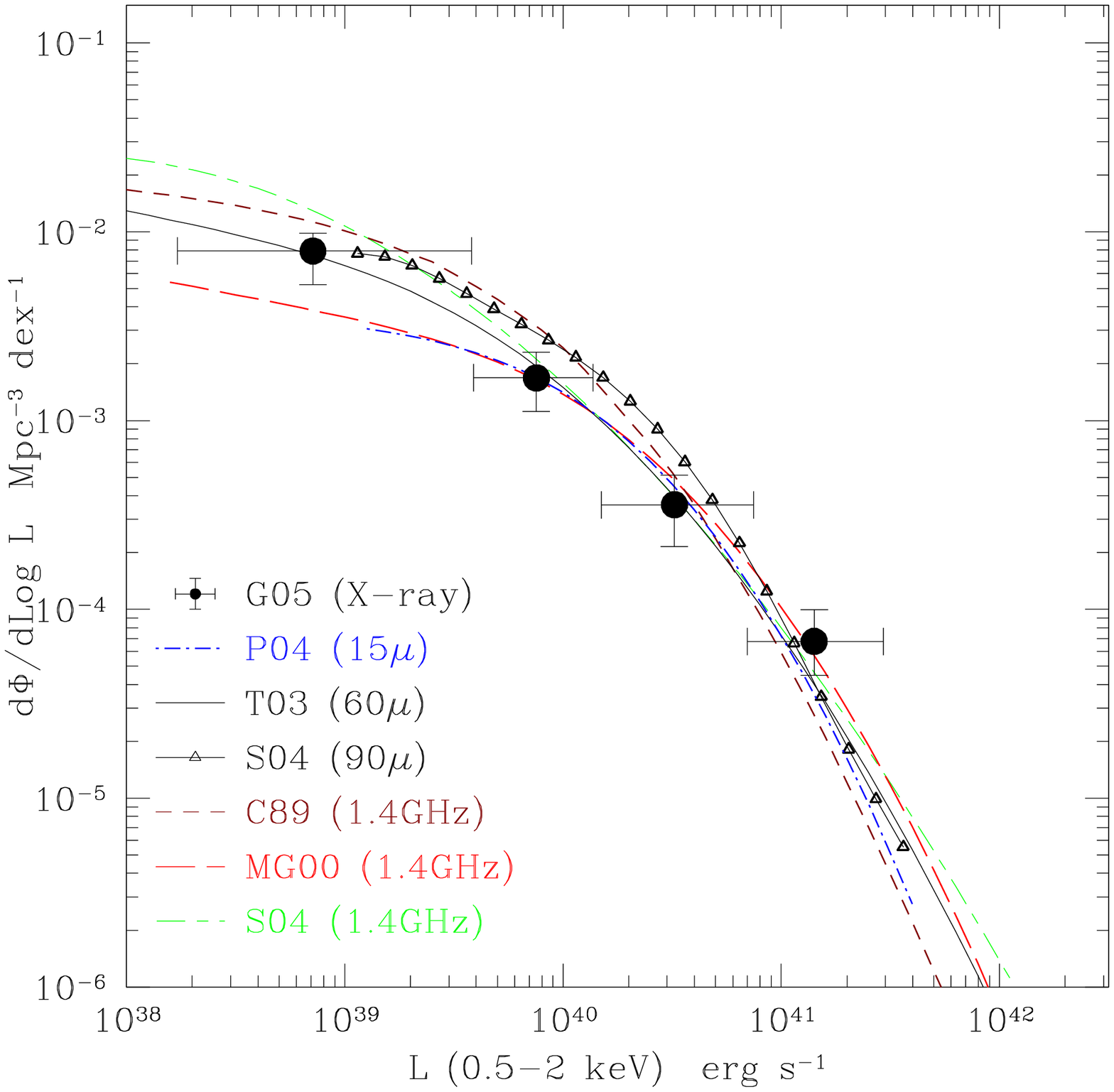}
   \includegraphics[width=.46\textwidth]{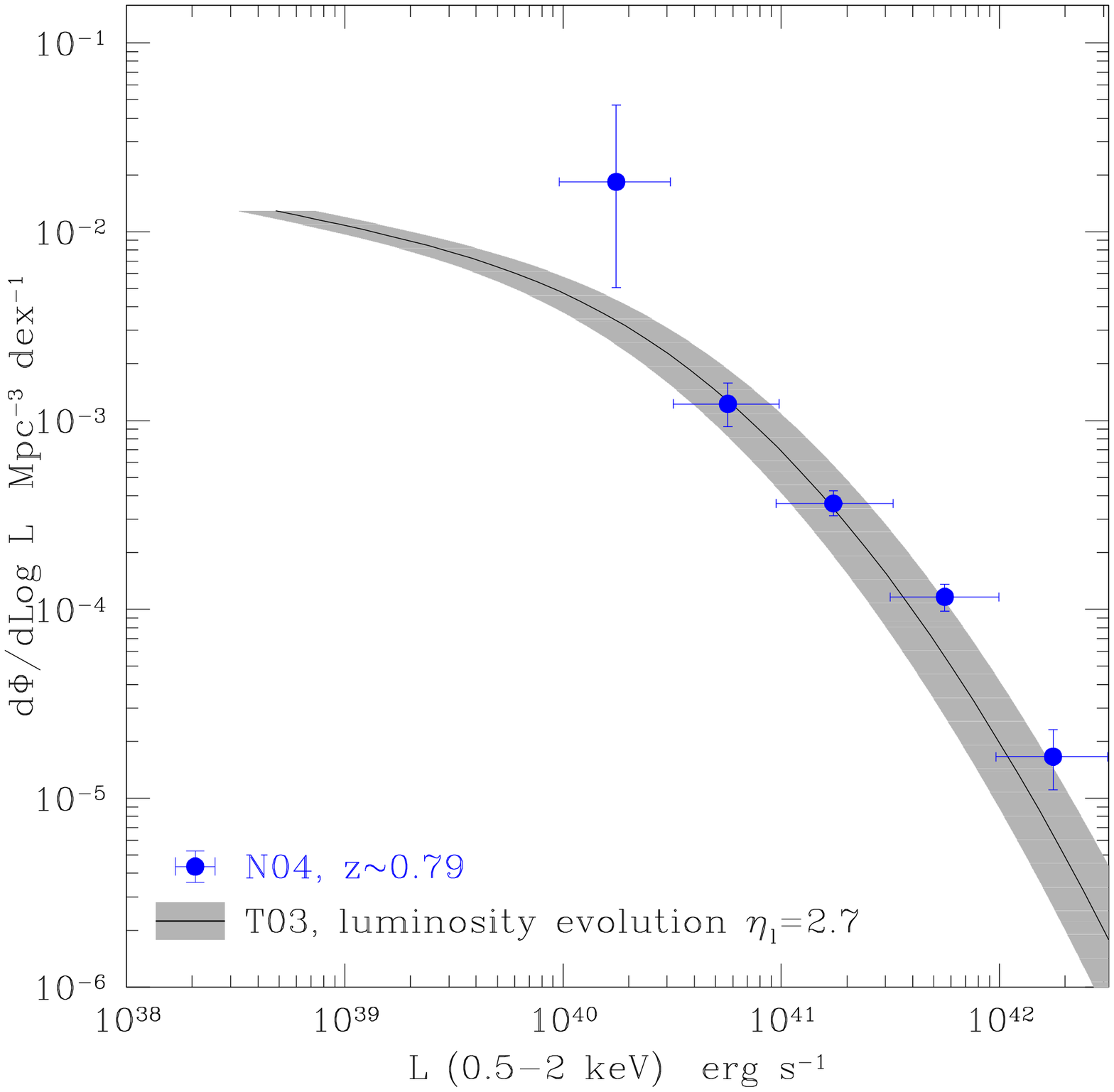}
   \caption { {\em Left:} IRAS, ISO, and radio local luminosity
     functions of SFG converted to the X-rays (see RCS05 for 
     references to the individual LFs). All the LFs converge to the
     same location. The large data points with error bars show an
     observational determination of the local XLF, based on \xmm\ data
     by \citet{georgantopoulos05}.
      \label{fi:xraytake} 
      {\em Right:} Comparison of the XLF derived from IRAS data (solid
      curve; the grey area shows the uncertainty on the evolution)
      with the XLF derived by \citet{norman04} in the \chandra\ Deep
      Field (data points with error bars).
      \label{fi:xlf-normanetal}
    }
\end{figure*}

The local differential luminosity function 
\begin{equation}
\varphi(\Log L)\, \de \Log L
\end{equation}
is defined as the comoving number density of sources per logarithmic
luminosity interval.  The evolution can be described
as pure luminosity with the form \citep{schmidt72}
\begin{equation}
L(z)\propto (1+z)^{\eta_{\rm l}}.     \label{eq:evolution}
\end{equation}

Infrared surveys provide a powerful method to select SFG, since the
bulk of the far and near IR emission is due to reprocessed light from
star formation, with AGN representing only a minor population
\citep{dejong84,franceschini01,elbaz02}.  The FIR LFs may be assumed
to be essentially unaffected by a contribution from Seyfert galaxies,
as the fraction of Seyferts is about $\sim 5$--10\% (RCS05).
While many determinations of IR LFs exist (see references in RCS05),
here we take \citet[][hereafter T03]{takeu03,takeu03err}
as reference. This is a 60$\mu$ LF derived from the IRAS Point
Source Catalog Redshift \citep{saun00} (PSC$z$). It includes 15,411
galaxies with $z\lesssim 0.07$, covering 84\% of the sky with a flux
limit of 0.6 mJy at $60\mu$.  While T03 reports pure-density evolution
for their LF, pure-luminosity may provide an equally good fit to the
data (T. Takeuchi, priv.\ comm.).

Other determinations of the SFG LF have been derived by the cross
correlation of radio surveys with optical ones (see references in RCS05).
The
redshifts covered in these surveys are similar to those of the T03
galaxies, but the number of objects is smaller due to a smaller sky
coverage, so reliable estimates of the evolution may not be derived.

\label{sec:xlf-discussion}

The local IR or radio LFs may be converted to X-ray ones by using the
approach first developed in \citet{avnitan86} (see also:
\citealt{ioannis99,norman04}), which may be summarised
as follows. Given a galaxy with IR or radio luminosity
$L$, let $P(L_{\rm X}|L)$ be the probability distribution of
the possible values of the galaxy's X-ray luminosity $L_{\rm X}$, as
given by the optical/IR/radio vs.\ X-ray correlations. Thus, the X-ray
LF may be otained by the convolution of an optical/IR/radio LF with
$P(L_{\rm X}|L)$.  In \citet{rcs03} it was reported that the X-ray
luminosity is tightly correlated with radio and FIR luminosities. By
assuming a Gaussian probability distribution for these correlations,
one has for example
\begin{eqnarray}
  P(\Log\ L_{\rm 0.5-2 keV}|\ \Log\ L_{60\mu}) = \nonumber\\
 =\frac{1}{\sqrt{2\pi}\sigma} \, {\rm e}^{-\frac{{\rm Log}\ 
    L_{60\mu} +9.05 - {\rm Log}\ \Lsx}{2\sigma^2}}
\end{eqnarray}
with $\sigma\sim 0.30$. 

A clear prediction for a $z=0$ XLF emerges from the comparison
of the infrared and radio LFs  (Fig.~\ref{fi:xraytake}, left
panel): the derived XLFs agree within a factor of 2 in the luminosity
interval $10^{40}$--$10^{41}$ \ergs, encompassing the knee region
after which all XLFs steepen toward higher luminosities; although
departures at lower and higher luminosities are present, the average
local X-ray luminosity density, $\sim (3\e{37} \pm 30\%)$ \ergs\ 
Mpc$^{-1}$, appears to be well defined.

\label{sec:xlfevol}

\citet{norman04} derived an XLF at higher redshifts (two bins: $\bar
z\sim 0.27$ and $\bar z\sim 0.79$; Fig.~\ref{fi:xlf-normanetal}, right
panel) than those probed by the IR and radio surveys discussed above.
Other strong constraints at high redshift come from the COMBO-17
survey \citep{wolf03}, and from the comparison of the observed X-ray
\lognlogs\ with that derived by integrating the XLF.  This work has
been done in detail in RCS05, and here we just quote the results: the
evolution is well described as pure-luminosity with an exponent
$\etalum\sim 2.7$, with possibly an hint that the evolution could be
stopped at $z\sim 1$ (Fig.~\ref{fi:xlf-normanetal}, right panel).

\section{Expected luminosity and redshift distributions with the WFXT}
\label{sec:wfxt}

The XLF derived in the previous section can be integrated in the volume
of space probed by the surveys to obtain
luminosity distributions
\begin{eqnarray}
   \frac{\de N}{\de\Log L}=\int_0^{z_\mathrm{max}} \!\!\!\!
   \varphi ( \Log L,z^\prime)\ \mathrm{min}[ V(z),   \nonumber\\
  V(\Log L,F_\mathrm{lim})] \ \de z
\end{eqnarray}
and redshift distributions
\begin{eqnarray}
   \label{eq:zdistr}
   \frac{\de N}{\de z}=\int_{\Log L_\mathrm{min}}^{\Log
     L_\mathrm{max}(z^\prime)}   \!\!\!\!
   \varphi(\Log L,z^\prime)\ \mathrm{min}[V(z),   \nonumber\\
  V(\Log L,F_\mathrm{lim})] \ \de \Log L
\end{eqnarray}
where $z^\prime=\mathrm{min}(z,z_\mathrm{stop})$;
$F_\mathrm{lim}$ is the limiting flux of the survey; $V(z)$ is the
comoving volume at redshift $z$; and $V(L,F)$ is the comoving volume
at the redshift at which a source with luminosity $L$ is observed with
flux $F$. All fluxes are considered in the 0.5--2.0 keV band.

For the following calculations, we take $z_\mathrm{max}=4$,
$z_\mathrm{stop}=1$, $L_\mathrm{min}=\mathrm{max}[10^{39}$, $4\pi
\mathrm{D}_\mathrm{lum}(z)^2 F_\mathrm{lim}]$ \ergs\ and
$L_\mathrm{max}=10^{42}(1+z^\prime)^{\eta_\mathrm{l}}$ \ergs. In
words, this means that we integrate on the luminosity range (at $z=0$)
$10^{39}$--$10^{42}$ \ergs, that we allow the maximum luminosity to
evolve with redshift, that we exclude luminosities lower than what
could visible given the redshift and limiting flux, and that the
integration is done up to $z=4$ but stopping the evolution at
$z_\mathrm{stop}=1$.  The evolution is pure-luminosity as in
Eq.~(\ref{eq:evolution}) with $\eta_\mathrm{l}=2.7$.

\begin{figure*}[tp]    
   \centering
   \includegraphics[width=.46\textwidth]{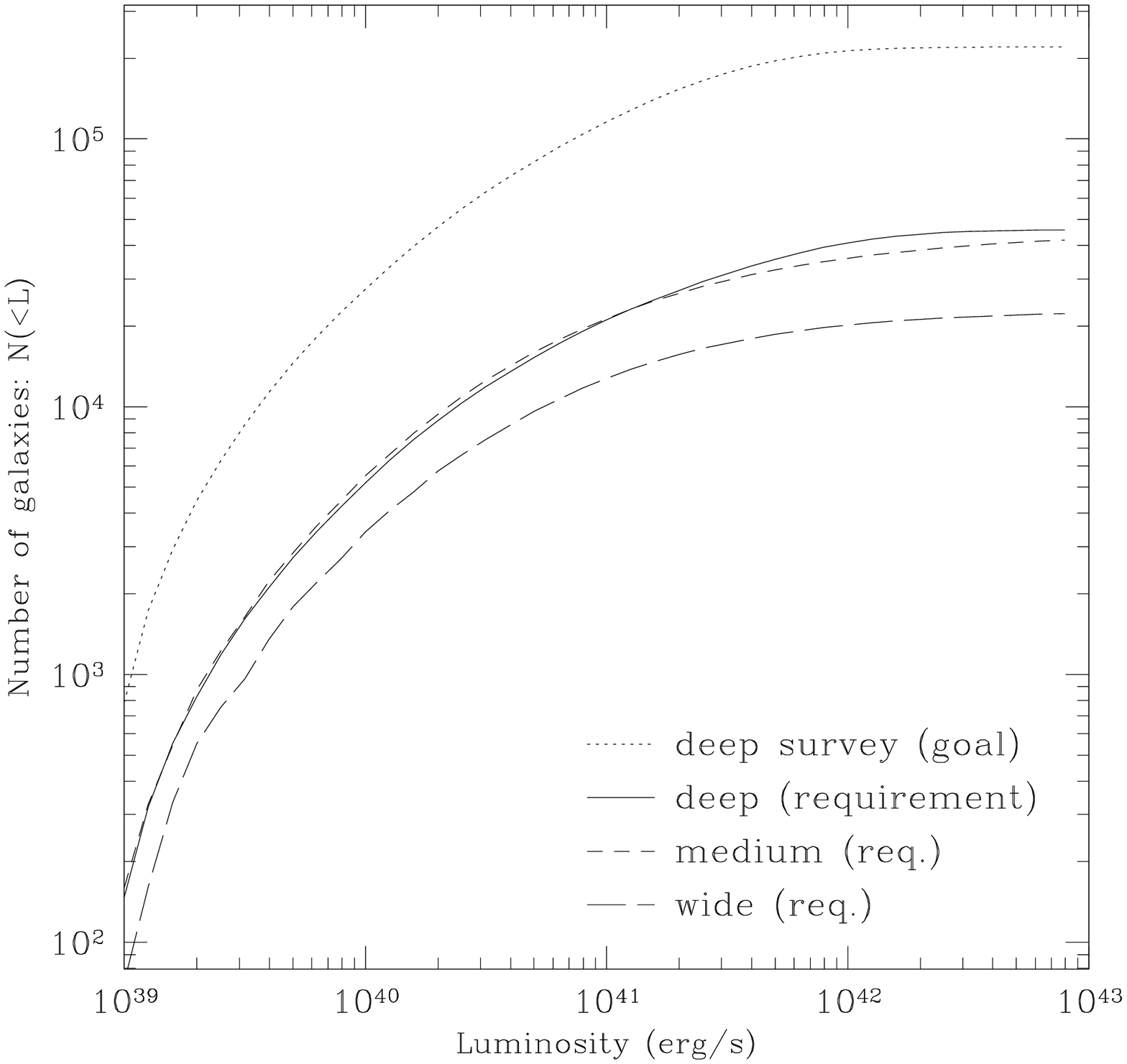}
   \includegraphics[width=.46\textwidth]{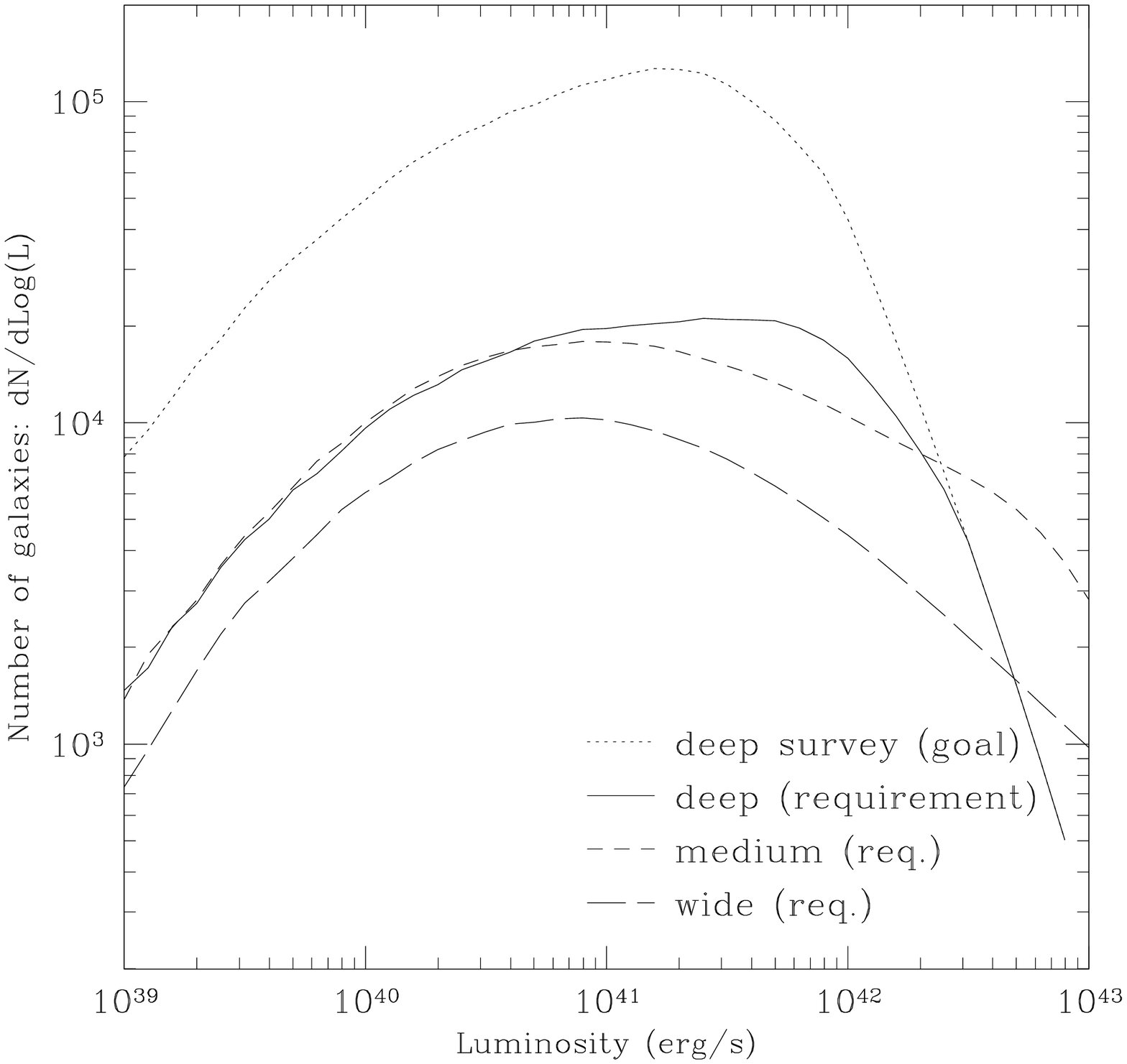}
    \caption { 
      {\em Left:} Expected cumulative luminosity distributions for SFG
      in the WFXT surveys.
      {\em Right:} Differential luminosity distributions. Since the
      knee of the SFG XLF is comprised (at $z=0$) in the range
      $10^{40}$--$10^{41}$ \ergs, it appears that the XLF will be well
      sampled by the WFXT.
}
\label{fi:lumdistr}
\end{figure*}

\begin{figure*}[tp]    
   \centering
   \includegraphics[width=.46\textwidth]{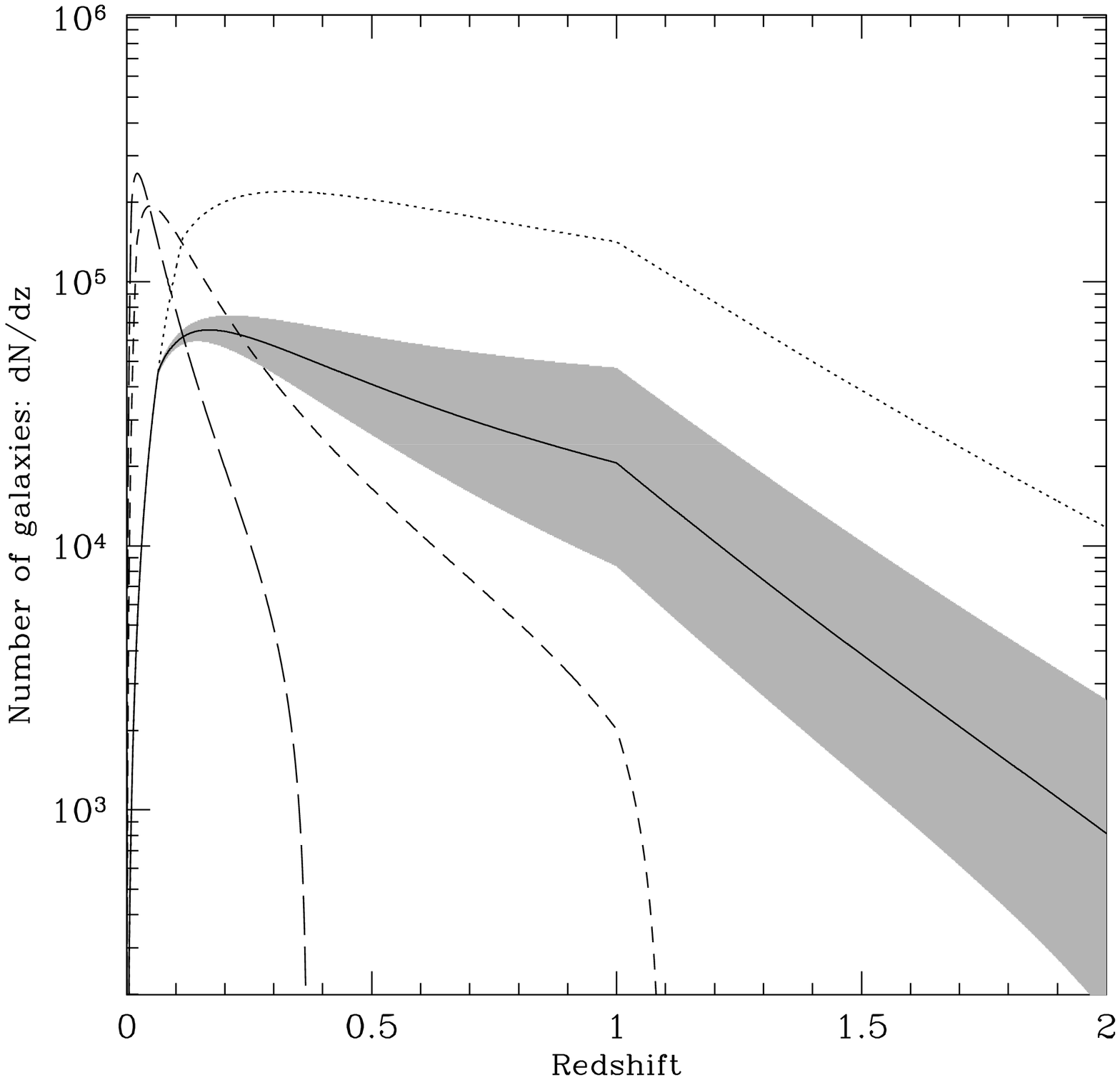}
   \includegraphics[width=.46\textwidth]{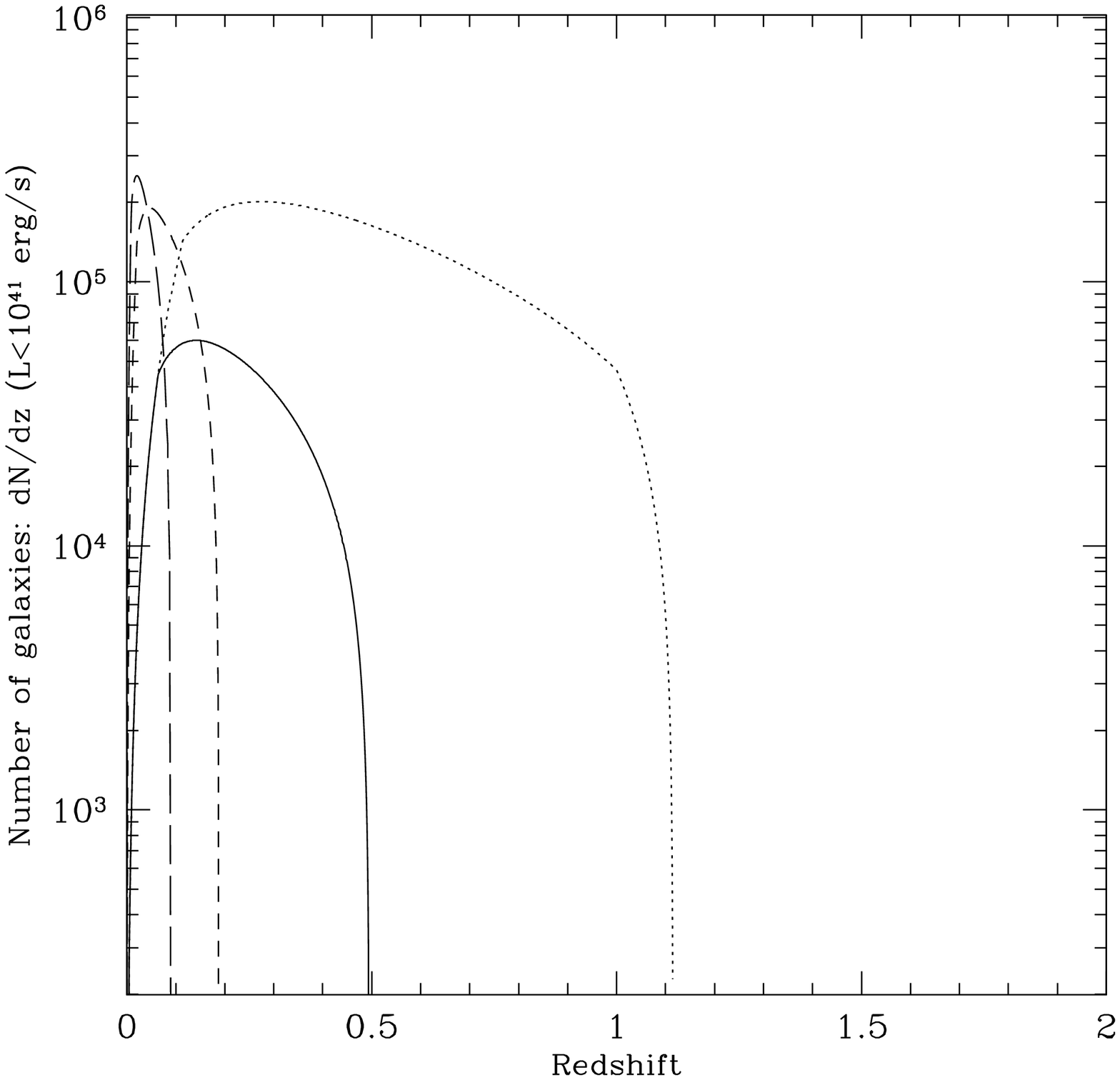}
   \caption { {\em Left:} Expected differential redshift distributions
     for SFG in the WFXT surveys. The grey area illustrates how the
     uncertainties about the XLF evolution could affect the \deep\
     survey.  {\em Right:} Same as left, but only considering galaxies
     with luminosity $L\le 3\e{40}(1+z^\prime)^\etalum$ \ergs: the
     knee region of the XLF will be probed up to $z\sim 0.5$--1.1.
     {\em Both:} Line styles as in Fig.~\ref{fi:lumdistr}.  }
\label{fi:zdistr}
\end{figure*}

The luminosity distribution is shown in Fig.~\ref{fi:lumdistr}, both
in cumulative (left panel) and differential form (right panel). The
cumulative form immediately shows the total number of SFG
which are expected to be detected in the WFXT surveys
($2\e4$--\,$4\e4$ objects per survey). Reaching the development goal
would enhance the number of SFG by a factor of $\sim\! 5$, up
to $2\e5$ objects in the \deep\ survey.

It is important to check that the SFG XLF will be well sampled at all
luminosities. From Fig.~\ref{fi:lumdistr} it is evident that at least
$10^3$ SFG with $L<2\e{39}$ \ergs\ should be detected in the \medium\
and \deep\ surveys, and that the ``knee'' region of the XLF (the range
$10^{40}$--$10^{41}$ \ergs\ at $z=0$, compare with Fig.~\ref{fi:xraytake}) will
be very well sampled with around $1.6\e4$ objects in each of the
\medium\ and \deep\ surveys.  Similarly, the high luminosity tail
($L>10^{42}$ \ergs) will also be well sampled with around $7\e3$
objects in the \medium\ survey. This part of the XLF is especially
important because objects in this luminosity range are quite rare, and
generally suspected of having a substantial part of their emission due
to an AGN.  Refined classification criteria, and the possibility of
doing spectral analysis will clearly be essential.

Reaching the development goal will enlarge the sample of SFG with
$L<10^{42}$ \ergs\ by a factor of $\sim 5$, while it should not make
much difference for brighter objects.

\smallskip
The expected redshift distribution is shown in Fig.~\ref{fi:zdistr}
(left panel). The \wide\ and \medium\ surveys should have redshift
peaks around $0.02$ and 0.05, respectively. Both will
provide sizable samples at larger redshift: $\sim 1000$ in the range
$0.2<z<0.3$ for the \wide\ survey, and $\sim 900$ in the range
$0.6<z<0.7$ for the \medium. The \deep\ survey will probe much higher
redshifts: $\sim 900$ objects with $1.2<z<1.3$, and other $\sim 900$
with $1.5<z<2.0$; these numbers would also be larger by a factor of
$\sim 5$, if the development goal is reached.  The
uncertainties on the SFG evolution are illustrated by the grey area in
Fig.~\ref{fi:zdistr} (left panel),
whose upper and lower edges correspond to evolution with $\etalum=3.4$
and 2.0, respectively.

However, it is likely that the highest redshift objects will also have
the larger luminosities. Thus it is reasonable to ask {\em up to which
redshift will the knee of the XLF be probed}. The local XLF
exhibits its knee in the range $10^{40}$--$10^{41}$ \ergs\
(Fig.~\ref{fi:xraytake}), thus we repeated the integration in
Eq.~(\ref{eq:zdistr}) taking $L_\mathrm{max}=3\e{40}(1+z^\prime)^\etalum$ \ergs.
 The result is shown in Fig.~\ref{fi:zdistr}
(right panel).  The \wide\ and \medium\ surveys will not probe the
knee of the XLF at redshift larger than $z\sim 0.1$ and $z\sim 0.2$,
respectively. The \deep\ survey will extend the probed redshift range
up to $z\sim\! 0.5$, while if the development goal is reached, redshifts
as large as $\sim\! 1.1$ could be observed.

\section{Discussion}
\label{sec:discussion}

From the expected luminosity and redshift distributions, it is evident
that the WFXT will be able to determine the SFG XLF with an accuracy
comparable to that of IRAS or optical surveys.  Thus there will be many new
possibilities to study how the X-ray emission depends on
other parameters, such as morphology, colours, redshift, etc. However,
such a work could only be made if multiwavelength information is
available. In fact, the first and most important task will be the
selection of the SFG, which are a minor fraction of the total of X-ray
surveys. Several different combinations of the same basic parameters
(X-ray luminosity, X-ray/optical flux ratio, hardness ratio, amount of
absorption, presence of broad lines in optical spectra, etc.) have
been explored by different authors in deep \chandra\ surveys (RCS05,
and references therein). All the determinations differ by up to a
factor of $\sim 2$; this scatter can be reduced only with a better
understanding of how these parameters are linked to each other, and
how they affect the selection (and the completeness of samples) of
SFG.  This only gets more difficult for wide-and-shallow surveys
(respect to deep pencil-beam ones) because the SFG/AGN fraction in
X-ray surveys depends on the limiting flux
(Fig.~\ref{fig:conteggi}). An attempt to investigate this problem for
a sample of SFG in the \chandra-COSMOS survey \citep{ccosmos-cat} may
be found in Ranalli et al.\ (2010, to be submitted). One of its main
results is that no rigid boundaries on the selection parameters can
be put; a sensible approach should build on statistical methods for
object classification.

The need for the most complete multiwavelength coverage also requires
that the choice of the sky areas covered by the WFXT surveys be
coordinated with (or follow, if not possible otherwise) other present
and future survey facilities (Pan-Starrs, the Large Synoptical Survey
Telescope, ALMA, LOFAR, E-VLA, etc.).

The planned WFXT surveys will be able to derive the SFG XLF and
determine its evolution with unprecedented accuracy up to $z\sim\!
0.5$ (1.1 if the development goals are reached) in the knee region,
and up to $z\sim\! 2$ (2.5) for the high-luminosity tail. Since the
cosmic star formation history as a peak in the range $1\lesssim
z\lesssim 2$, it is evident that the goals should be
pursued with strong commitment.  
The cosmic accretion history has a peak at a similar redshift, and the
two phenomena seem to have shared a very similar trend. Thus, the
larger the probed redshift range, the more impact the WFXT 
will have for studies of SFG and AGN coevolution.

Finally, were the goals reached, and the numbers still on the safe
side of the confusion limit, some ultra-deep pointings should be
considered as very profitable. E.g., observing an area of 10 deg$^2$ with a
limiting flux of $10^{-17}$ \ergscmq\ would extend the coverage of the
knee of the XLF up to $z\sim 1.7$, and of the high-luminosity tail up to
redshifts well beyond the peak of the cosmic star formation history.

\begin{acknowledgements}
We thank Roberto Gilli and Andrea Comastri for stimulating discussions.
\end{acknowledgements}

\bibliographystyle{aa}
\bibliography{../fullbiblio}

\end{document}